\documentclass{ws-ijmpe}
\usepackage{xcolor}
\usepackage[super,compress]{cite}
\usepackage[verbose]{hyperref}
\hypersetup{colorlinks=false,allbordercolors=blue,pdfborderstyle={/S/U/W 1}}
\usepackage{url}
\usepackage{ulem}

\usepackage{xcolor}

\usepackage{xcolor}

\usepackage[dvipsnames]{xcolor}

\begin{document}

\thispagestyle{plain}

\def\bib{B\kern-.05em{I}\kern-.025em{B}\kern-.08em}
\def\btex{B\kern-.05em{I}\kern-.025em{B}\kern-.08em\TeX}

\title{Charm quark energy loss fluctuations in clumpy color string medium}

\author{Daria Prokhorova}

\address{Department of Physics, Tsinghua University, Haidian District\\
Beijing, 100084,
China\\
daria-prokhorova@mail.tsinghua.edu.cn}

\author{Shuzhe Shi}

\address{Department of Physics, Tsinghua University, Haidian District\\
Beijing, 100084,
China\\
shuzhe-shi@tsinghua.edu.cn}

\author{Evgeny Andronov}

\address{Physics Department, St. Petersburg State University, Peterhof\\
St. Petersburg, 198504,
Russia\\
e.v.andronov@spbu.ru}

\maketitle


\begin{abstract} 
We simulate charm quark propagation through an ensemble of color flux tubes formed between receding color charges in proton-proton collisions. Under the assumption of sharp string boundaries and a localized nature of the overlap regions, high flux tube densities lead to the formation of string clusters. We convert the local energy density of the superposed color field into an effective temperature of an ideal gluon gas and study the dependence of charm energy loss on its spatial inhomogeneities. By comparing different energy-density extraction prescriptions, we show that both elastic and inelastic momentum damping mechanisms are sensitive to the medium properties. However, the collisional channel is an order of magnitude smaller in absolute terms (up to $\sim0.15$~GeV) compared to the radiative loss (up to $\sim1.5$~GeV). Overall, the results demonstrate that the spatial overlap of individual strings generates a complex local energy density pattern, which predominantly governs the medium-induced quenching of charm quarks.
\end{abstract}



\section{Introduction}

The dynamics of inelastic high-energy proton-proton (p+p) collisions can be described via the formation of multiple color flux tubes between receding color charges~\cite{Capella:1992yb,Werner:1993uh}. These longitudinally extended fields, finite in the transverse dimension, form so-called color strings that effectively capture the confinement features of non-perturbative QCD~\cite{Artru:1974hr}. Within this picture, color strings either decay to hadrons via the Schwinger mechanism~\cite{Schwinger:1951nm} or act as a pre-equilibrated non-perturbative medium~\cite{Fries:2005yc}. The latter can be modeled, for example, via string melting to partons, as in the AMPT framework~\cite{Lin:2004en}, the dynamical formation of a hydrodynamically evolving core in DCCI2~\cite{Kanakubo:2021qcw} or the core-corona separation in the EPOS4 generator~\cite{Werner:2023zvo,Werner:2023jps}. Similarly, in the PHSD microscopic transport approach~\cite{Cassing:2009vt}, pre-hadrons melt into partonic quasi-particles in regions where local energy density exceeds a critical threshold. In the color rope model~\cite{Biro:1984cf}, dense field regions are described as overlapping sources forming a $K$-fold rope. In the Color String Percolation Model~\cite{Braun:2015eoa}, they overlap and form string clusters~\cite{Braun:1991dg} that acquire modified superposed color fields under the assumption of sharp boundaries of the overlapping string disks. This introduces the spatial redistribution of energy initially available for medium formation and creates clumpy structures within the system~\cite{Prokhorova:2026ziv}. We follow this string fusion approach, which we have previously applied to study the correlations and fluctuations of bulk observables~\cite{Andronov:2023vnh, Prokhorova:2023hgq, Prokhorova:2024ext}. 

In this work, we perform a thermal matching of the local energy density within the color string volumes to that of an ideal gluon gas. As in Ref.~\cite{Fries:2005yc}, we omit the detailed kinetic description of the medium formation and do not impose any EOS~\cite{Prokhorova:2025jmd}, focusing solely on the extraction of the local chromo-electric field energy density $\varepsilon$. Following Ref.~\cite{Prokhorova:2026ziv}, we test $\varepsilon \propto k$ and $\varepsilon \propto \sqrt{k}$ scaling prescriptions on the integer number of locally overlapped strings $k$. They yield two distinct effective local temperature profiles, corresponding to the temperatures the gluon plasma with the same energy density would have had if it thermalized without transverse expansion. To resolve the resulting spatial fluctuations of $T$ on an event-by-event basis, we estimate the energy loss of heavy-flavor (HF) quarks, produced in the primary hard scatterings of the same event, as they propagate through the string ensemble.

There exist various models of hard-soft interactions at different stages of the system evolution, ranging from solving the Fokker-Planck equations in Glasma fields~\cite{Carrington:2022bnv,Carrington:2021dvw} to describing the Brownian motion of HF quarks in QGP~\cite{Svetitsky:1987gq,Cao:2018ews,2306.07966}. Common to all these approaches is that, being much heavier than other scales in the problem, HF quarks have significantly longer relaxation times than light partons, preventing their full thermalization within the lifetime of the QCD environment~\cite{He:2022ywp}. Consequently, they serve as excellent tomographic probes of the medium.

In our model, charm quarks undergo both elastic and inelastic scatterings off the gluons in both the halo and the hot spots of the effective string medium. While the collisional momentum damping with the Thoma–Gyulassy elastic cross-section~\cite{Thoma:1990fm, Peigne:2008nd} was previously estimated in this framework~\cite{Prokhorova:2025jmd}, we introduce the radiative energy loss channel~\cite{Djordjevic:2003zk,Djordjevic:2007at} in the present study. Specifically, we employ the Djordjevic–Gyulassy–Levai–Vitev (DGLV) formalism~\cite{Gyulassy:1999zd,Djordjevic:2003zk} to describe medium-induced gluon bremsstrahlung. Here, we focus on the isotropic thermal limit, leaving the momentum-anisotropic case~\cite{Romatschke:2003ms,Romatschke:2004jh} for future work. Given the strong temperature dependence of both loss mechanisms, we also determine the event-by-event energy-density-weighted eccentricity of the string medium, which provides a key initial condition for anisotropic hydrodynamic models.

The paper is organized as follows. Section~\ref{sec2} presents our framework, while results and conclusions are given in Section~\ref{sec3} and Section~\ref{sec4}, respectively.

\section{Model framework}
\label{sec2}

In this work, we extend our hybrid Monte-Carlo event generator developed in Ref.~\cite{Prokhorova:2025jmd} to describe both elastic and inelastic interactions of charm quarks traversing a short-lived medium formed in p+p collisions at LHC energies.

\subsection{Hard sector}

In each event, we assume that a charm $c\bar{c}$ pair is produced in a primary hard scattering. Both quarks are assigned the same transverse production point, which is sampled in the string-occupied regions according to the local string density. This separation of $c\bar{c}$ production from the underlying event is similar to the approach of Ref.~\cite{Singh:2025duj}, where PYTHIA-generated quarks are placed at binary vertices for Langevin propagation in the medium. Likewise, Ref.~\cite{Beraudo:2023nlq} generates charm pairs with POWHEG-BOX~\cite{Alioli:2010xd} interfaced with PYTHIA~\cite{Bierlich:2022pfr} and embeds them in a hydrodynamic background, but samples positions by entropy rather than by string density. In fully integrated frameworks, the HF quarks are treated as part of the DGLAP evolution of the multi-parton scattering, as in EPOS4HQ~\cite{Zhao:2023ucp}, where they can be produced in space- or time-like cascades or in the Born process. 

We limit the study to charm quarks with transverse momentum $2 \leq p_T \leq 20$~GeV and rapidity $|y| \leq 0.5$. Within these kinematic constraints, charm quarks are generated with PYTHIA 8.3~\cite{Bierlich:2022pfr} via the $g+g \rightarrow c\bar{c}$ and $q+\bar{q} \rightarrow c\bar{c}$ subprocesses, with initial-state radiation (ISR) enabled and final-state radiation (FSR) disabled. The ISR is retained to preserve the realistic momentum distribution
of the incoming partons, while the FSR is switched off to study the in-medium evolution of charm quarks. At leading order, the $c\bar{c}$ pair is emitted back-to-back in the transverse plane with an isotropic orientation of the pair axis. While the recoil from ISR breaks this exact correlation, affecting pair-level observables of the initial state, the azimuthal distribution of individual quarks remains isotropic. Furthermore, we neglect the $c\bar{c}$ potential, assuming it is effectively screened by the medium. While the dissociation dynamics of bound states in such environments are investigated in other studies~\cite{Zheng:2026alo}, we treat the $c$ and $\bar{c}$ as independent probes.

Given the large charm quark mass compared to the relevant momentum scales, we neglect the virtuality of the produced quarks and treat them as on-shell particles immediately after the hard scattering. This approach is similar to the on-shell transport stage of the EPOS4HQ framework~\cite{Zhao:2023ucp,Zhao:2024oma}, but differs from the JETSCAPE setup~\cite{Putschke:2019yrg}, where high-virtuality probes are explicitly evolved through a medium-modified vacuum shower by the MATTER module~\cite{Majumder:2013re,Cao:2017qpx} before passing the resulting low-virtuality partons to the on-shell LBT~\cite{He:2015pra,Cao:2016gvr}, MARTINI~\cite{Schenke:2009gb}, or CUJET~\cite{Xu:2014ica,Xu:2015bbz} transport codes. The charm hadronization is left for another study, so that all the observables presented below are quark-level quantities.

\subsection{Soft sector}

We model the initial energy deposition in a high-energy proton-proton collision within the Gribov-Regge formalism~\cite{Gribov:1962fw,Abramovsky:1973fm}, where multiple pomeron exchanges encode the multi-parton dynamics. Derived from the analytical properties of the S-matrix, this effective approach is supported by the topological expansion of QCD in the large-$N_c$ limit~\cite{tHooft:1973alw}. In turn, the unitarity cut of the dominant cylindrical pomeron diagram yields two color strings~\cite{Artru:1974hr, Werner:1993uh}. They are attached to the constituent or sea partons within receding protons, whose longitudinal momenta (proton momentum fractions) determine the rapidity positions of the strings~\cite{Shen:2017bsr, Prokhorova:2025jmd}. The interaction between the partons at the string endpoints can be described by the Cornell potential, which includes a linear confinement term proportional to the string tension $\sigma = 0.9$~GeV/fm $\approx 0.18$~GeV$^2$, in addition to the Coulomb contribution. We neglect the latter at the relevant long-range scales of the problem. Moreover, these longitudinally extended objects are assumed to have a finite transverse size, thus forming color flux tubes rather than purely one-dimensional objects. A parallel can be drawn with the Glasma fields~\cite{Gelis:2012ri} generated immediately after a collision, whose characteristic transverse size is naturally determined by the saturation scale of the Color Glass Condensate~\cite{Gelis:2010nm}. However, this analogy is purely qualitative, as color strings are phenomenological objects, and their dynamics is not derived from classical Yang-Mills equations.

With triple-pomeron vertices neglected, the multi-pomeron structure of the interaction yields a substantial string density even in p+p collisions at high energies. Moreover, the finite transverse extent of strings ($r_0 = 0.25$~fm) leads to a rich structure in the string density distribution arising from diverse overlap configurations. To account for its granular structure, we evaluate the chromo-electric field energy density in small volume elements (referred to as cells) using either \(\varepsilon_k = \sqrt{k}\,\varepsilon_1\) or \(\varepsilon_k = k\,\varepsilon_1\), where \(k\) is the integer local string overlap order in a given cell and $\varepsilon_1 = \sigma/(\pi r^2_0)$ is the energy density of a single string. The former corresponds to scaling with the chromo-electric field strength, while the latter follows the field energy density~\cite{Prokhorova:2026ziv}. 

We map the local energy density to an effective temperature by treating the energy in each cell as originating from an ideal gluon gas. Specifically, we match the cell energy density $\varepsilon_\mathrm{cell}$ from the geometrical string overlaps to the thermodynamic relation, yielding $T_\mathrm{cell} = \sqrt[4]{30\varepsilon_\mathrm{cell}/(16\pi^2)}$. This neither implies a microscopic string-to-gluon conversion nor imposes equality of the complete field and kinetic energy-momentum tensors, including the transverse and longitudinal pressures.

In this approach, the effective medium inherits screening properties~\cite{Gossiaux:2008jv,Gossiaux:2009hr} defined by the local gluon Debye mass $\mu_\mathrm{cell}(T_\mathrm{cell})$ that is computed by solving the self-consistent equation $\mu_\mathrm{cell}(T_\mathrm{cell}) = T_\mathrm{cell}\sqrt{4\pi\alpha_s(\mu_\mathrm{cell}^2(T_\mathrm{cell}))}$, where $\alpha_s(\mu^2(T))$~\cite{Xu:2014tda,Shi:2018izg} is the thermal running coupling. For the rest of the paper, we omit the $_\mathrm{cell}$ subscript, keeping all medium quantities local. We note that in a momentum-anisotropic medium both the effective temperature $T$~\cite{Prokhorova:2025jmd} and the functional dependence $\mu(T)$ are modified~\cite{Strickland:2011aa,Dong:2021gnb}, a detailed investigation of which we leave for a future study.

\subsection{Hard-soft interactions}

We model the propagation of charm quarks through the effective medium. Assuming the medium remains static during a single interaction, we freeze the gluon ensemble at each time step $\Delta t = 0.1$ fm and propagate the quarks up to $t_\mathrm{max} = 1.5$ fm.

\subsubsection{Elastic interactions}
For collisional energy loss, we implement the elastic $2 \rightarrow 2$ scattering off medium gluons using the infrared-finite Thoma--Gyulassy cross-section~\cite{Thoma:1990fm,Peigne:2008nd}
\begin{equation}
\frac{\mathrm{d}\sigma}{\mathrm{d}\hat{t}} = \frac{2\pi \alpha_s^{2}(\hat{t})}{\left(\hat{t}+\mu^2(\hat{t})\right)\hat{t}},
\end{equation}
where $\hat{t}$ is the Mandelstam variable for momentum transfer. As derived in our previous work~\cite{Prokhorova:2025jmd}, the momentum loss rate for a finite-mass HF quark scattering off gluons in an ideal gas is given by
\begin{equation}\label{eq.loss_ideal_bose_gluons}
\begin{split}
\frac{\mathrm{d}p}{\mathrm{d}t}
&= \frac{2\pi}{3} \alpha_s(\mu^2) \, \alpha_s(6ET)
\ln\!\Bigg(
\frac{6ET}{\mu^2} - \frac{m^2}{\mu^2}
+ \frac{m^4}{12pT\mu^2}
\ln\left(\frac{m^2 +6T(E+p)}
{m^2 +6T(E-p)}\right)\Bigg) \\
&\times T^2 \frac{1-\beta}{\beta^2}\Bigg[\frac{1}{1-\beta} - \operatorname{arctanh}\beta\cdot\left(1+\frac{2}{\beta}\right)+ 1 \\
&+ \frac{12\zeta(3)}{\pi^2}\frac{T}{E}\cdot\left(\frac{\operatorname{arctanh}\beta}{\beta} - 1 \right )\Bigg],
\end{split}
\end{equation}
where $E = \sqrt{p^2 + m^2}$ is the quark energy and $\beta = p/E$. To characterize the opacity of the medium, we also track the average number of elastic scatterings accumulated by the quark along its trajectory. The corresponding scattering rate, following the CUJET3 framework~\cite{Xu:2014tda}, is given by:
\begin{equation}\label{eq:ncoll}
    \frac{dN_\mathrm{col}}{dt} =  \frac{4\alpha_s(\mu^2)\alpha_s(6ET)}{\mu^2} \frac{18\zeta(3)}{\pi}T^3.
\end{equation}

In the simulation, these rates are evaluated cell-by-cell using the appropriate local $T$. Summing the local loss and the local $N_\mathrm{col}$ up to \(t_\mathrm{max} = 1.5\)~fm yields the total energy loss along the quark path and the average number of scatterings, respectively. Here, the collisional loss is implemented as a deterministic momentum decrement along a straight trajectory, neglecting stochastic momentum kicks and transverse broadening, which is adequate for our single-particle observables.

\subsubsection{Inelastic interactions}
The inelastic momentum loss of a HF quark propagating through the medium is dominated by medium-induced gluon bremsstrahlung~\cite{Gyulassy:1993hr}. Recently, the AMY (Arnold-Moore-Yaffe)~\cite{Arnold:2001ba,Arnold:2001ms,Arnold:2002ja} and DGLV (Djordjevic-Gyulassy-Levai-Vitev)~\cite{Gyulassy:1999zd,Djordjevic:2003zk} formulations of radiative energy loss were rigorously compared within the JETSCAPE Monte-Carlo framework~\cite{Shi:2022rja}. In this work, we adopt the DGLV formalism, which builds upon the earlier GLV approach~\cite{Gyulassy:2000er} and is based on the opacity expansion~\cite{Wiedemann:2000za} in a finite-size dynamic plasma~\cite{Djordjevic:2008iz,Djordjevic:2007at}. The single-gluon emission rate at first order is
\begin{equation}\label{eq.dglv}
\begin{split}
z\frac{d\Gamma^\mathrm{n=1}_{q\to qg}}{dz} (p, T, t, z)  &= \frac{6}{\pi^2} \,\rho(T) \int d^2\mathbf{k}_\perp \int d^2\mathbf{q}_\perp \Bigg\{\frac{\alpha^2_s\left[\mathbf{q}^2_\perp\right]}{\mathbf{q}^2_\perp(\mathbf{q}^2_\perp + \mu^2)} \\ 
&\times \alpha_s\left[\frac{\mathbf{k}^2_\perp}{z_+(1-z_+)}\right] \cdot \frac{z}{z_+} \cdot \left|\frac{dz_+}{dz}\right| \frac{-2}{(\mathbf{k}_\perp - \mathbf{q}_\perp)^2 +\chi^2}\\
& \times \left( \frac{\mathbf{k}_\perp \cdot (\mathbf{k}_\perp-\mathbf{q}_\perp)}{\mathbf{k}^2_\perp + \chi^2} - \frac{(\mathbf{k}_\perp-\mathbf{q}_\perp)^2}{(\mathbf{k}_\perp-\mathbf{q}_\perp)^2 + \chi^2}\right)\\
& \times \left(1 - \cos{\left[\frac{(\mathbf{k}_\perp-\mathbf{q}_\perp)^2 + \chi^2}{2z_+p}\cdot t\right]}\right)\Bigg\},
\end{split}
\end{equation}
where $\rho(T)$ is the local medium gluon density, \(p\) is the initial momentum of the HF quark, \(T\) is the local effective medium temperature, \(z\) is the fractional momentum carried by the emitted gluon, and \(t\) is the time since the last splitting. The integrals in Eq.~\eqref{eq.dglv} run over emitted gluon transverse momentum $|\mathbf{k}_\perp| < zp$ and the momentum transfer $|\mathbf{q}_\perp| < \sqrt{6pT}$. The light-cone momentum fraction $z_+$ is related to $z$ by $z_+ = \frac{1}{2}\left(z + (z^2 - \left(\mathbf{k}_\perp/p\right)^2)^{1/2}\right)$. The regulator $\chi^2(T) = m^2 z_+^2 + \frac{\mu^2(T)}{2} (1-z_+)$ controls the soft collinear divergences and reflects the dead-cone effect~\cite{Dokshitzer:1991fc} for the massive quark ($m = 1.5$ GeV for charm). Finally, the cosine term encodes the Landau-Pomeranchuk-Migdal (LPM) interference effect~\cite{Migdal:1956tc}, which suppresses emission shortly after a previous splitting.

For each medium cell traversed by the charm quark, we compute the mean number of emitted gluons \(\langle N_\mathrm{rad}\rangle = \int dz \int dt \, [d\Gamma/dz(p,T,t,z)]\). The emission probability in Poisson form is then $P_\mathrm{em} = 1 - \exp(-\langle N_\mathrm{rad}\rangle)$, which in the limit of a sufficiently small time step $P_\mathrm{em} \approx \langle N_\mathrm{rad}\rangle$. In case \(\langle N_\mathrm{rad}\rangle\) exceeds unity, multiple emissions are possible, but they are strongly suppressed by the LPM effect. Therefore, the formation time is reset after each emission~\cite{Shi:2022rja}. If an emission occurs according to $P_\mathrm{em}$, the momentum fraction $z$ is sampled from the corresponding $dN/dz = \int dt \, [d\Gamma/dz(p,T,t,z)]$ distribution, the quark momentum is reduced by $\Delta p = z p$ and the quark energy is recalculated to restore the mass shell. Here, the transverse momentum of the radiated gluon and the recoil momentum transferred to the medium are neglected in the quark kinematics. 

To enable efficient sampling during the simulation, the DGLV spectrum of Eq.~\eqref{eq.dglv}, fully integrated over the transverse momenta $\mathbf{k}_\perp$ and $\mathbf{q}_\perp$, is precomputed on a four-dimensional grid of $p$, $T$, $t$, and $z$ values. At runtime, the rate at arbitrary parameter values is obtained via spline interpolation of this grid. The details of the numerical integration and the tabulation procedure will be presented elsewhere.

\section{Results}
\label{sec3}
In this section, we present the results for charm quark transport through the string medium. We focus on the sensitivity of both collisional (blue) and radiative (red) momentum quenching mechanisms to the two energy-density extraction schemes. 

\subsection{Average momentum loss}
\label{subsec:avg}

\begin{figure}[h!]
\centering
\includegraphics[width=0.9\linewidth]{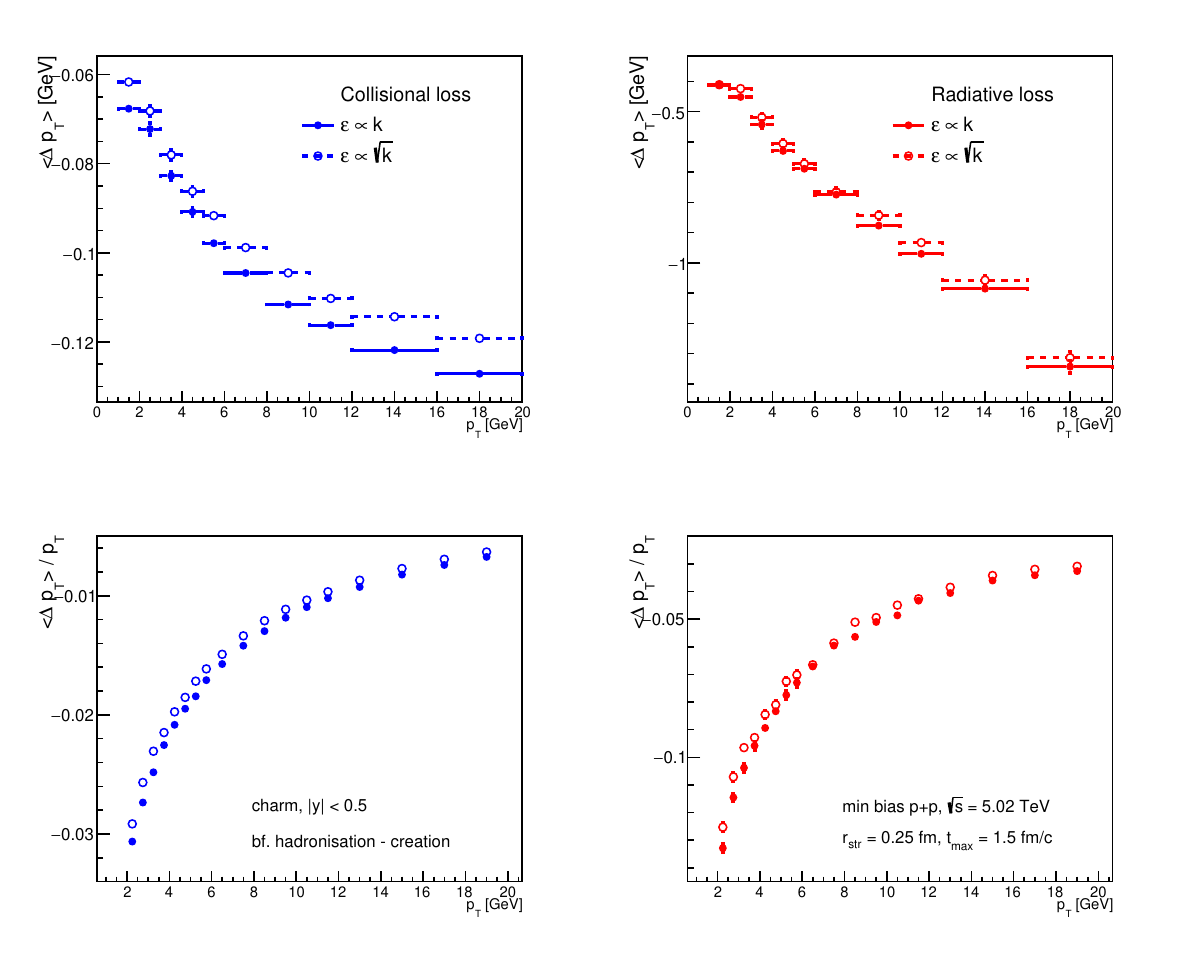}
\caption{Average momentum loss (top row) and relative momentum loss (bottom row) of charm quarks as functions of the initial $p_T$ for collisional (left column) and radiative (right column) energy loss. Full (empty) circles correspond to $\varepsilon \propto k$ ($\varepsilon \propto \sqrt{k}$) scaling for the thermal isotropic gluon medium.}
\label{fig:absloss}
\end{figure}

Fig.~\ref{fig:absloss} shows the average momentum loss $\langle \Delta p_T \rangle$ (top row) and the relative momentum loss $\langle \Delta p_T \rangle/p_T$ (bottom row) as functions of the initial charm $p_T$ for collisional (left column) and radiative (right column) channels in the ideal gluon gas. In all cases, the $k$-scaling (full circles) yields slightly stronger quenching than the $\sqrt{k}$-scaling (empty circles). Radiative loss dominates, exceeding the collisional one by an order of magnitude at $p_T = 20$~GeV.

\begin{figure}[h]
\centering
\includegraphics[width=0.9\linewidth]{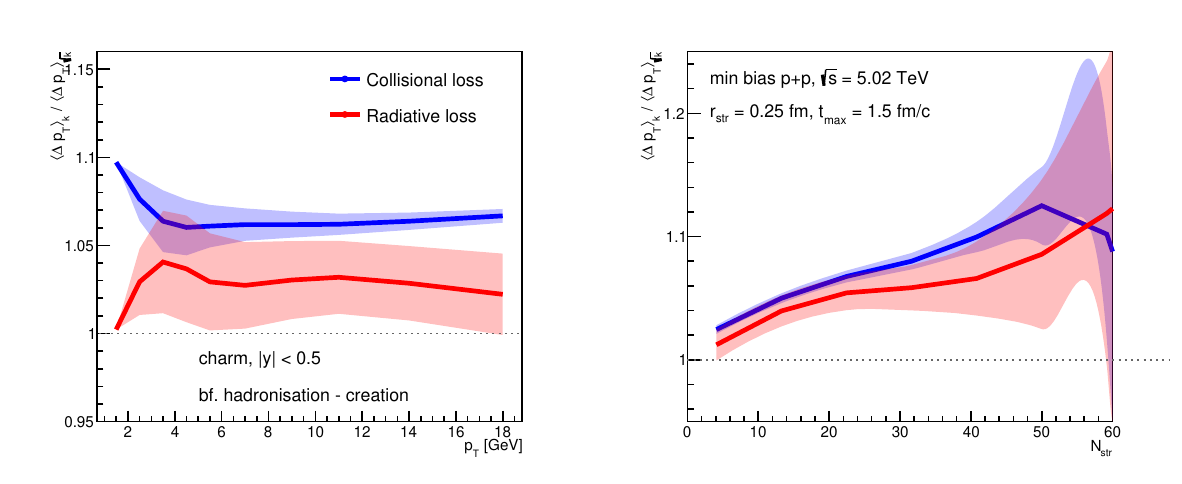}
\caption{Ratio of the average momentum loss for $\varepsilon \propto k$ and $\varepsilon \propto \sqrt{k}$ as a function of initial charm $p_T$ (left) and number of strings in event $N_\mathrm{str}$ (right) for collisional (blue) and radiative (red) quenching mechanisms in thermal isotropic gluon medium.}
\label{fig:ratio}
\end{figure}

Fig.~\ref{fig:ratio} shows the ratio $R = \langle \Delta p_T \rangle_{k} / \langle \Delta p_T \rangle_{\sqrt{k}}$ for both energy-loss mechanisms in the thermal isotropic medium as a function of the initial charm transverse momentum $p_T$ (left) and as
a function of the number of strings $N_\mathrm{str}$ (right). While the ratio is relatively insensitive to the quark's initial momentum, its strong dependence on $N_\mathrm{str}$ (which acts as a proxy for event activity) implies that the distinction between the two energy-density schemes becomes much more pronounced in high-multiplicity events.

\subsection{Event-by-event fluctuations}
\label{subsec:fluct}

Since the string medium is intrinsically inhomogeneous, the energy loss strongly depends on the specific path and production point of the quark. Combined with the event-by-event variation of the medium geometry, this results in broad fluctuations of the quenching. 

\begin{figure}[h]
\centering
\includegraphics[width=\linewidth]{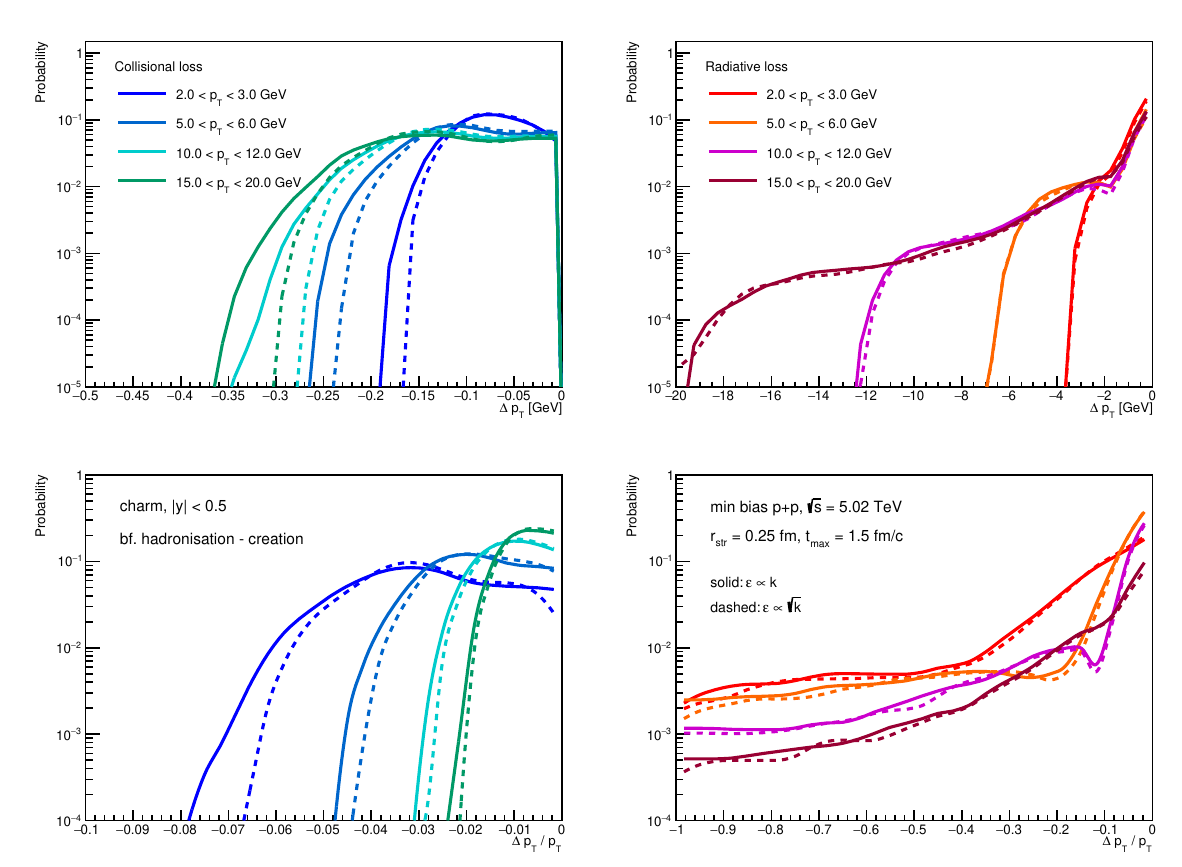}
\caption{Probability distribution of the momentum loss $\Delta p_T$ (top row) and relative momentum loss $\Delta p_T/p_T$ (bottom row) for four $p_T$ intervals (colors), shown for collisional (left column) and radiative (right column) quenching mechanisms for momentum-isotropic medium with $\varepsilon \propto k$ (solid curves) or $\varepsilon \propto \sqrt{k}$ (dashed curves) scaling.}
\label{fig:distributions}
\end{figure}

Fig.~\ref{fig:distributions} shows the probability distributions of the absolute \(\Delta p_T\) (top row) and relative \(\Delta p_T/p_T\) (bottom row) momentum loss for four initial charm-quark $p_T$ intervals (colors). The collisional distributions (left column) are peaked at small losses and widen smoothly with $p_T$. The radiative distributions (right column) instead exhibit long power-law tails extending down to $|\Delta p_T| \simeq 20$~GeV, reflecting the stochastic nature of gluon bremsstrahlung. The $k$-scaling (solid curves) systematically produces broader distributions and more extended tails compared to the $\sqrt{k}$-scaling (dashed curves) in both energy-loss channels, highlighting the stronger hot-spot structure. 

\begin{figure}[h]
\centering
\includegraphics[width=0.9\linewidth]{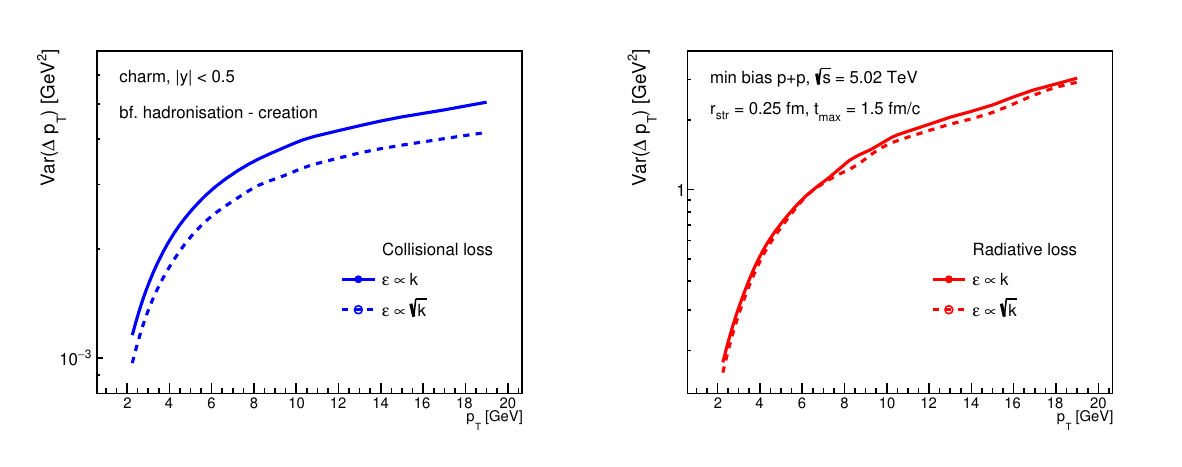}
\caption{Variance of the momentum loss as a
function of initial charm $p_T$ for collisional (left) and radiative (right) quenching mechanisms, in the isotropic gluon medium with $\varepsilon \propto k$ (solid) or $\varepsilon \propto \sqrt{k}$ (dashed) scalings.}
\label{fig:variance}
\end{figure}

The width of these distributions is quantified in
Fig.~\ref{fig:variance}, where the variance
$\mathrm{Var}(\Delta p_T)$ is plotted as a function of initial charm $p_T$. For radiative loss, it is roughly two orders of magnitude larger than for collisional loss, and both grow similarly with $p_T$. While the $k$-scaling systematically enhances the variance of both channels, the effect is much more pronounced for collisional loss.

\subsection{Medium geometry and temperature fluctuations}
\label{subsec:ecc}

\begin{figure}[h]
\centering
\includegraphics[width=0.9\linewidth]{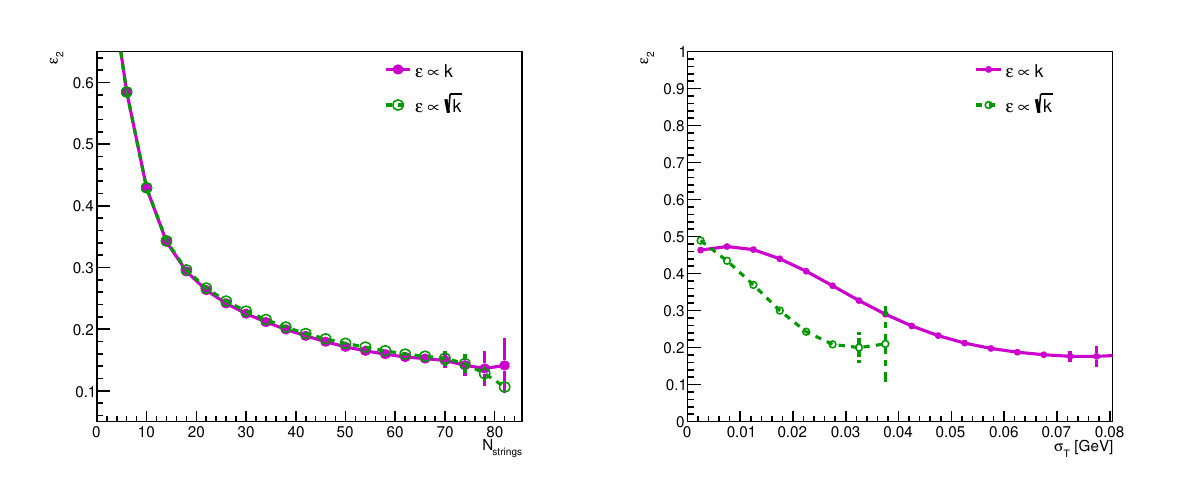}
\caption{Energy-density-weighted initial eccentricity $\varepsilon_2$ at midrapidity as a function of the number of strings in event $N_\mathrm{str}$ (left) and of the event-averaged temperature standard deviation $\sigma_T$ (right), for $\varepsilon \propto k$ (full) and
$\varepsilon \propto \sqrt{k}$ (empty) energy density scalings.}
\label{fig:eccentricity}
\end{figure}

Finally, we examine the influence of event geometry on charm-quark momentum loss. The left panel of Fig.~\ref{fig:eccentricity} shows the initial energy-density-weighted eccentricity $\varepsilon_2$ at midrapidity versus the number of strings in event, $N_\mathrm{str}$. As the latter increases, the string overlap region becomes denser and more symmetric, reducing $\varepsilon_2$ from $\simeq 0.6$ (at $N_\mathrm{str} \simeq 6$) to $\simeq 0.15$ (at $N_\mathrm{str} \simeq 70$), with the curves for $k$- and $\sqrt{k}$-scalings slightly diverging at high $N_\mathrm{str}$. The right panel further reveals that at fixed $\varepsilon_2$ the $k$-scaling yields larger temperature fluctuations, $\sigma_T$, than the $\sqrt{k}$-scaling, reflecting the non-trivial superposition of chromo-electric fields in the string clusters.

\begin{figure}[h]
\centering
\includegraphics[width=0.9\linewidth]{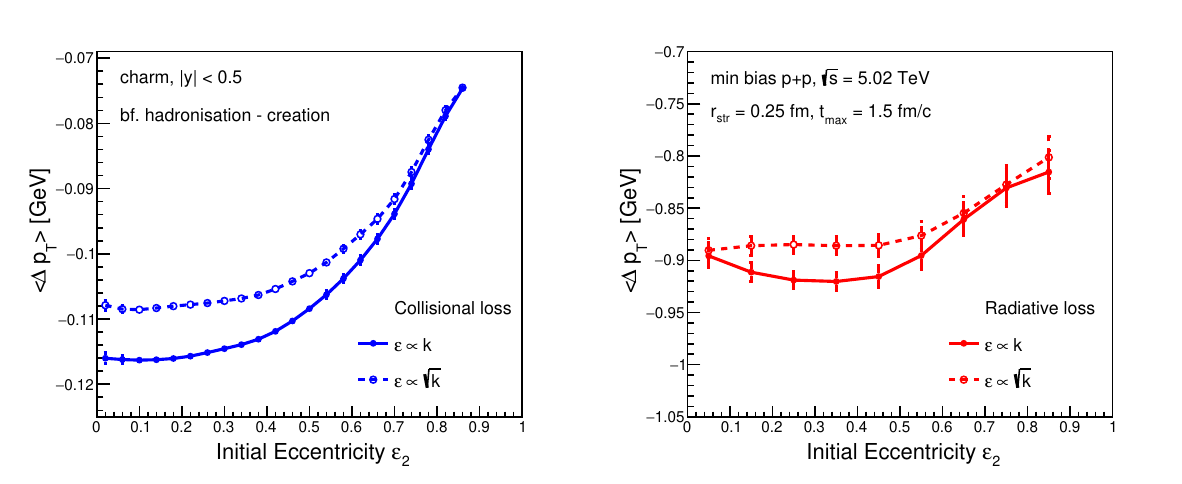}
\caption{Average charm quark momentum loss as a function of the local eccentricity $\varepsilon_2$ for collisional (left) and radiative (right) quenching mechanisms, in the isotropic gluon medium with $\varepsilon \propto k$ (full) or $\varepsilon \propto \sqrt{k}$ (empty) scalings.} 
\label{fig:loss_vs_ecc}
\end{figure}

Fig.~\ref{fig:loss_vs_ecc} directly correlates the average energy loss with the event geometry, plotting it against the energy-density-weighted eccentricity $\varepsilon_2$ at the quark rapidity. For collisional loss (left), quarks in denser and more symmetric regions (lower $\varepsilon_2$) lose more momentum. In contrast, the radiative loss (right) exhibits a weaker relative variation with eccentricity, revealing only a mild anti-correlation of $\langle p_T \rangle$ with $\varepsilon_2$.

\begin{figure}[h]
\centering
\includegraphics[width=\linewidth]{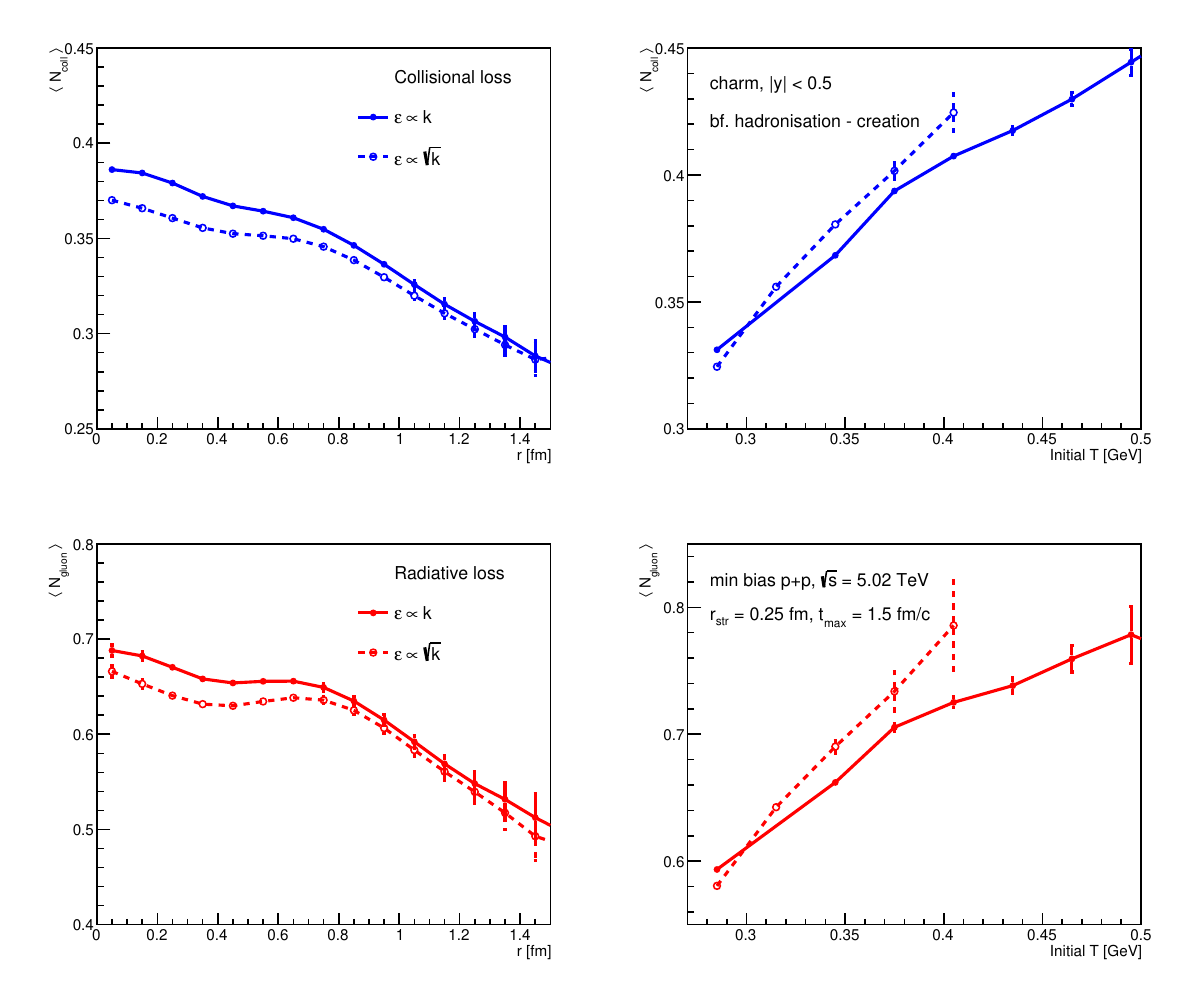}
\caption{Average number of elastic scatterings \(\langle N_\mathrm{col} \rangle\) (top row) and radiated gluons \(\langle N_\mathrm{gluon} \rangle\) (bottom row) versus the radial distance from the medium center \(r\) (left column) and the local temperature \(T\) (right column), evaluated at the start of the quark trajectory in the isotropic gluon medium with $\varepsilon \propto k$ (full) or $\varepsilon \propto \sqrt{k}$ (empty) scalings.}
\label{fig:interactions}
\end{figure}

\begin{figure}[h]
\centering
\includegraphics[width=\linewidth]{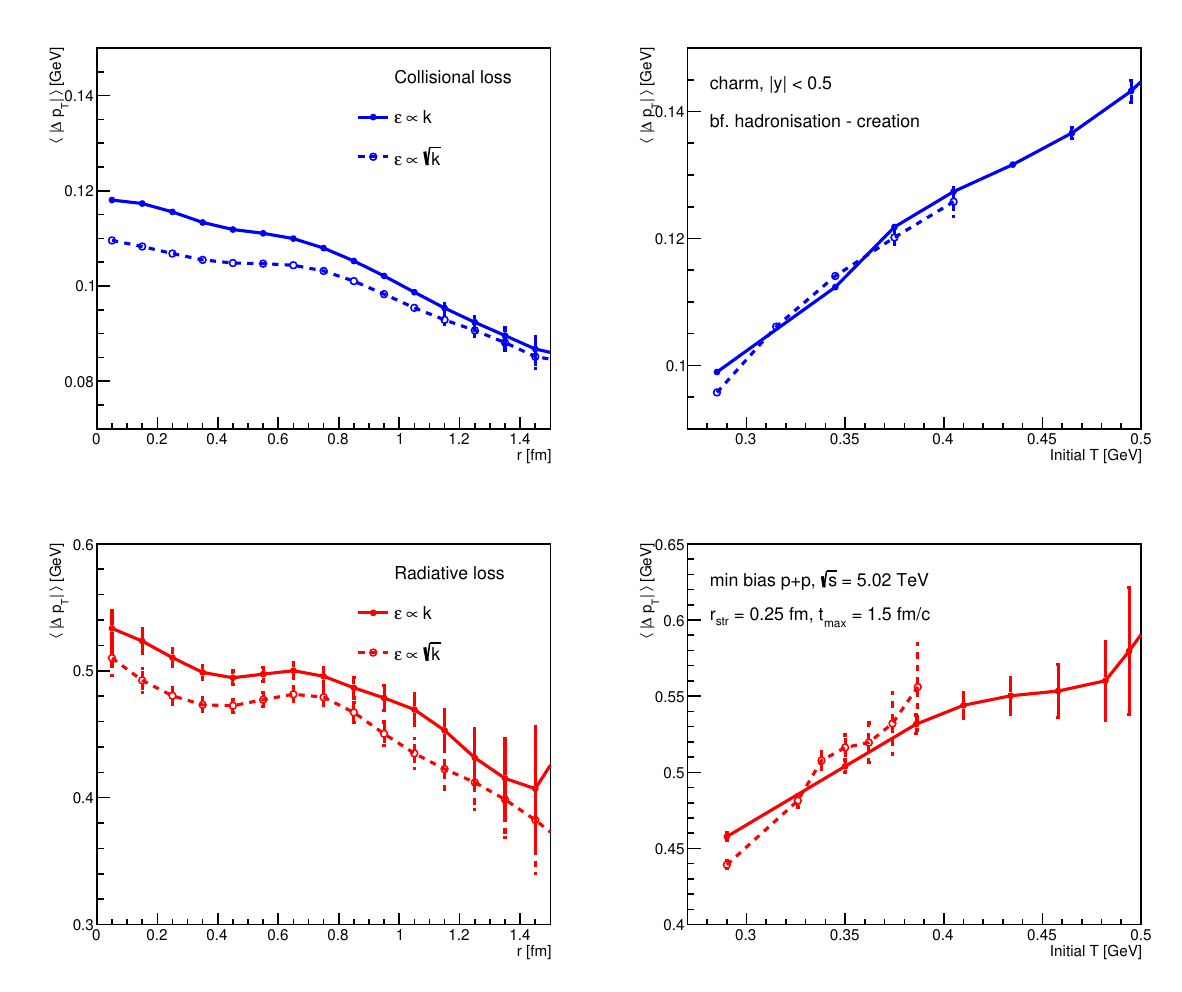}
\caption{Average absolute momentum loss \(\langle |\Delta p_T| \rangle\) for collisional (top row) and radiative (bottom row) quenching mechanisms versus the radial distance from the medium center \(r\) (left column) and the local temperature \(T\) (right column), evaluated at the start of the quark trajectory in the isotropic gluon medium with $\varepsilon \propto k$ (full) or $\varepsilon \propto \sqrt{k}$ (empty) scalings.}
\label{fig:loss_geometry}
\end{figure}

\begin{figure}[h]
\centering
\includegraphics[width=\linewidth]{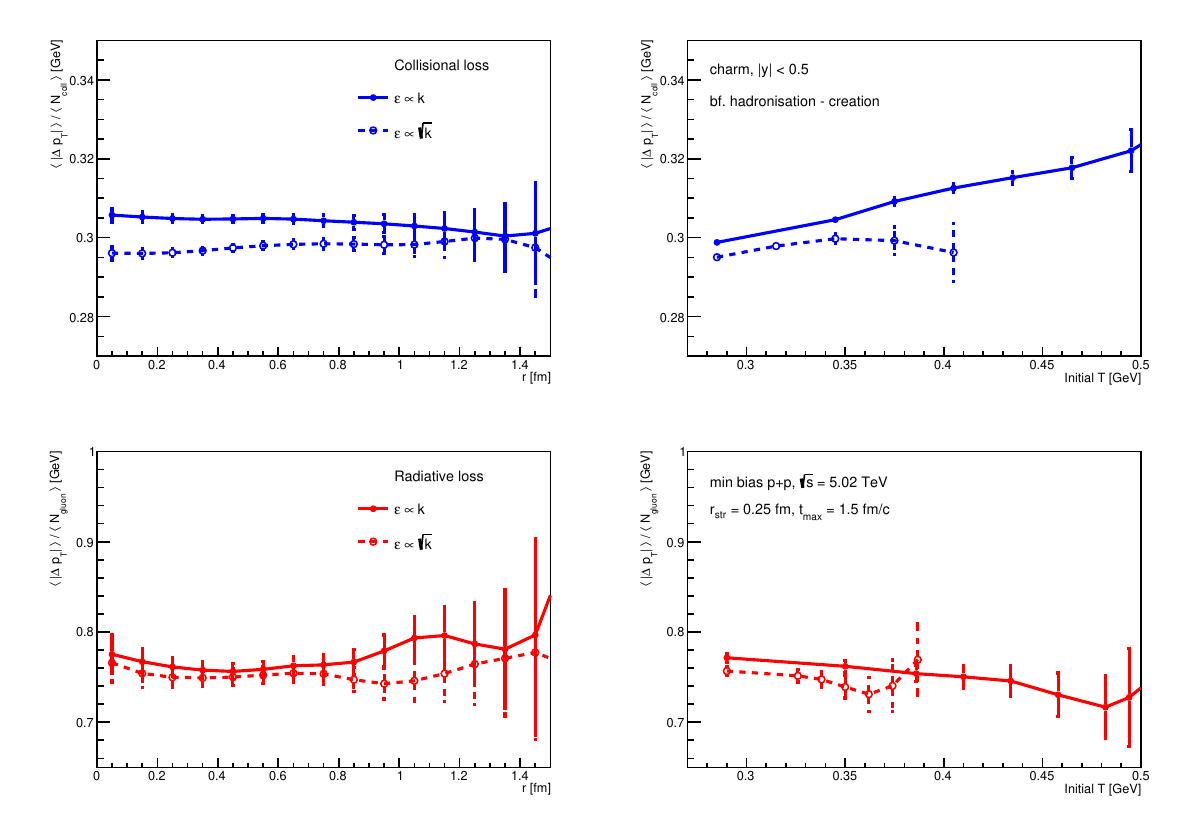}
\caption{Effective absolute momentum loss per interaction \(\langle |\Delta p_T| \rangle / \langle N_\mathrm{col} \rangle\) for collisional (top row) and \(\langle |\Delta p_T| \rangle / \langle N_\mathrm{gluon} \rangle\) for radiative (bottom row) quenching mechanisms, as a function of the radial distance from the medium center \(r\) (left column) and the local initial temperature \(T\) (right column), evaluated at the start of the quark trajectory in the isotropic gluon medium with $\varepsilon \propto k$ (full) or $\varepsilon \propto \sqrt{k}$ (empty) scalings.}
\label{fig:loss_per_interaction}
\end{figure}

Fig.~\ref{fig:interactions} shows the dependence of the average number of elastic scatterings \(\langle N_\mathrm{col} \rangle\) the quark undergoes (top row) and the average number of gluons \(\langle N_\mathrm{gluon} \rangle\) it radiates (bottom row) on the medium conditions at the start of its trajectory. Specifically, we examine how its distance from the energy-density-weighted medium center (left column) and the local effective temperature (right column) affect its subsequent transport. Both counts generally decrease with increasing radial distance and increase with the initial temperature. The resulting momentum loss is shown in Fig.~\ref{fig:loss_geometry} for the collisional (top row) and radiative (bottom row) channels. The quenching is also maximized if the quark starts propagation closer to the medium center ($r\rightarrow0$) and at higher temperatures, which reflects the dense core and dilute halo structure. The common trend suggests measuring effective collisional and radiative losses per interaction. In Fig.~\ref{fig:loss_per_interaction} we observe that they remain approximately constant with radial distance and initial temperature. Consequently, the momentum loss is predominantly determined by the change in the accumulated number of collisions and radiated gluons, rather than by a local change in the energy transferred per single interaction.

\section{Conclusions}
\label{sec4}
In this work, we extended our hybrid Monte-Carlo event generator to study the propagation of charm quarks through the inhomogeneous color string medium formed in p+p collisions at $\sqrt{s} = 5.02$~TeV. By implementing both elastic and inelastic scatterings, we investigated how the spatial energy-density fluctuations affect the quenching observables.

We find that while the mean collisional loss is relatively sensitive to the energy-density extraction scheme ($\varepsilon \propto k$ versus $\varepsilon \propto \sqrt{k}$), the radiative channel dominates the total quenching by an order of magnitude and dictates the overall theoretical uncertainty. The linear $k$-scaling amplifies the hot-spot structure, leading to a hotter effective medium and significantly broader event-by-event fluctuations. 

Our results demonstrate that momentum loss in small p+p systems is not negligible and should be accounted for in jet-quenching studies. Furthermore, its strong sensitivity to the medium properties provides valuable constraints on the initial state. While the present analysis is restricted to the isotropic medium, extending our framework to the momentum-anisotropic case will provide further input to anisotropic hydrodynamics.
A follow-up study including heavy-quark hadronization and final-state observables is currently in preparation.

\section*{Acknowledgments}

DP acknowledges support of Shui Mu scholarship of Tsinghua University. SS is supported by the National Key Research and
Development Program of China under Contract No. 2024YFA1610700, by the NSFC under grant No. 12575143, and by Tsinghua University under grant Nos. 04200500123, 531205006, 533305009.

\bibliographystyle{ws-ijmpe}
\bibliography{sample}
\end{document}